\documentclass[superscriptaddress,amsmath,amssymb,aps,prb,showkeys,preprint]{revtex4-2}

\usepackage{graphicx}%
\usepackage{dcolumn}%
\usepackage[english]{babel}
\usepackage{bm}%
\usepackage{tabularx}%
\usepackage{enumerate}
\usepackage[colorlinks, citecolor=blue, linkcolor=blue]{hyperref}
\usepackage{multirow}%
\usepackage{amsmath,amssymb,amsfonts}%
\usepackage{amsthm}%
\usepackage{mathrsfs}%
\usepackage[title]{appendix}%
\usepackage{xcolor}%
\usepackage{textcomp}%
\usepackage{booktabs}%
\usepackage{algorithm}%
\usepackage{algorithmicx}%
\usepackage{algpseudocode}%
\usepackage{listings}%

\begin{document}

\title{Topological meron-antimeron domain walls and skyrmions in a low-symmetry system}

\author{Reiner~Br\"uning} \email{rbruenin@physnet.uni-hamburg.de}
\affiliation{Institute of Nanostructure and Solid State Physics, University of Hamburg, Jungiusstrasse 11, 20355 Hamburg, Germany}

\author{Levente~R\'{o}zsa} \email{rozsa.levente@wigner.hun-ren.hu}
\affiliation{Department of Theoretical Solid State Physics, HUN-REN Wigner Research Centre for Physics, H-1525 Budapest, Hungary}
\affiliation{Department of Theoretical Physics, Budapest University of Technology and Economics, H-1111 Budapest, Hungary}

\author{Roberto~Lo~Conte}
\affiliation{Institute of Nanostructure and Solid State Physics, University of Hamburg, Jungiusstrasse 11, 20355 Hamburg, Germany}
\affiliation{Zernike Institute for Advanced Materials, University of Groningen, 9747 AG Groningen, The Netherlands}

\author{André~Kubetzka}
\affiliation{Institute of Nanostructure and Solid State Physics, University of Hamburg, Jungiusstrasse 11, 20355 Hamburg, Germany}

\author{Roland~Wiesendanger}
\affiliation{Institute of Nanostructure and Solid State Physics, University of Hamburg, Jungiusstrasse 11, 20355 Hamburg, Germany}

\author{Kirsten~von~Bergmann}
\affiliation{Institute of Nanostructure and Solid State Physics, University of Hamburg, Jungiusstrasse 11, 20355 Hamburg, Germany}

\date{\today}

\begin{abstract}

The generation of topologically non-trivial magnetic configurations has been a pivotal topic in both basic and applied nanomagnetism research. Localized non-coplanar magnetic defects such as skyrmions or merons were found to interact strongly with currents, making them interesting candidates for future spintronics applications. Here, we study a low-symmetry bcc(110) system by spin-polarized scanning tunneling microscopy and an atomistic spin model using parameters obtained from first-principles calculations. We demonstrate how a delicate balance between energy terms generates both topologically trivial and non-trivial domain walls, depending on their crystallographic direction. The topological walls consist of merons and antimerons and the topological charge amounts to about 0.2/nm wall length. The incorporation of holes in the films facilitates the transition from an in-plane ferromagnetic ground state to a spin-spiral state. Both 
domain walls and spirals transition into isolated elongated magnetic skyrmions in applied magnetic fields, establishing low-symmetry systems as a versatile platform for spin-texture engineering.


\end{abstract}

\maketitle
\section*{Introduction}\label{sec1}
\vspace*{0.3\baselineskip}


Topological magnetic defects have attracted considerable research attention over the past decades~\cite{Nagaosa2013, GoebelPR2021}. The topology of magnetic textures is governed by the relative local orientation of the magnetic moments or spins mapped onto the unit sphere, with all spins pointing along the same direction corresponding to the topologically trivial ferromagnetic state. 
When the spins locally deviate from the ferromagnetic direction, several different spin textures can arise. 
Vortices or merons~\cite{PhysRevB.91.224407,Yu2018, li_stabilizing_2014, gao_creation_2019, gao_spontaneous_2020, Amin2023, Bhukta2024-rr, maranzana_merons_2024, xu_spontaneous_2024}, where the spins cover half of the unit sphere, are found in in-plane magnetized films. 
Skyrmions~\cite{Bogdanov1989, RommingS2013, Leonov_2016, Jiang_2015, Woo2016, MoreauLuchaire2016, LoConteNANO-LETT2020}, where the spins cover the whole unit sphere, typically arise as a knot in the magnetization within an out-of-plane magnetized background. When similar knots are present in in-plane magnetized films, they are referred to as bimerons~\cite{Zhang2015,Heo2016,Goebel2019}. All these spin structures are non-coplanar and locally possess a finite topological charge. 
Exploring the spin texture of these structures has been one of the main driving forces behind the rapid development of magnetic imaging techniques and the adaptation of field-theoretical concepts from particle physics to  
magnetic systems. Such 
topologically non-trivial 
magnetic defects are stable against perturbations, making them promising candidates for storing data, where their creation, annihilation and motion is strongly influenced by the details of their local spin textures~\cite{Fert2013,Tomasello2014,Zhang2015}.

Magnetic domain walls in ferromagnets are extended defects separating 
domains with different spin orientations. Domain walls may be distinguished based on the rotational 
direction of the spins inside them. In Bloch walls, the spins rotate in the plane perpendicular to the normal vector of the wall, while in Néel walls some spins  
align parallel to the normal vector. The spins are 
coplanar in both Bloch and Néel walls. 
In contrast, cross tie walls~\cite{10.1063/1.1723105} are non-coplanar structures consisting of an alternating chain of 
vortices and antivortices. 
The dipole-dipole interaction plays a decisive role in domain formation in thin films with thicknesses ranging from tens of nanometers to micrometers, and it may prefer Bloch, Néel or cross tie walls depending on the layer thickness~\cite{10.1063/1.1729367} and sample geometry~\cite{Wiese_2007,li_stabilizing_2014}. However, the long-range nature of the dipolar interaction prevents downscaling these walls to the nanometer regime where local magnetic interactions become dominant. The most studied among these in the context of domain walls is the Dzyaloshinsky-Moriya interaction (DMI) present in non-centrosymmetric systems~\cite{dzyaloshinsky1958thermodynamic,Moriya1960}, which promotes domain-wall formation with a preferred rotational sense of the spins, 
which can be either Bloch-type or Néel-type  
depending on the symmetry of the crystal. 
Inside these coplanar domain walls merons and skyrmions only exist as localized defects, 
as has been predicted theoretically~\cite{PhysRevB.99.184412,amari_PhysRevB.109.104426, chen_magnetic_2024} and observed experimentally~\cite{Bhukta2024-rr,Amin2023}. A domain wall in the form of a chain of merons and antimerons, analogous to cross tie walls, has only been reported in centrosymmetric systems without 
DMI~\cite{gao_spontaneous_2020,Lv2022} or in such non-centrosymmetric systems where the dipolar interaction competes with the 
DMI~\cite{Nagase2021}.

Spin spirals are extended coplanar spin structures where the spins rotate along the direction of the modulation vector $\boldsymbol{q}$. The domain walls 
between domains of spirals with different orientations of $\boldsymbol{q}$ vectors and thus different rotational planes can consist of a chain of merons~\cite{PhysRevLett.108.107203,Yu2010,Schoenherr2018,BrueningPRB2022,maranzana_merons_2024}, with their spacing depending on the  
orientation of the wall with respect to the $\boldsymbol{q}$ vectors inside the domains. 

While most studies have focused on high-symmetry systems such as cubic crystals or thin films with at least a threefold rotational symmetry around the normal vector of the film, modulated spin structures have also been discovered in low-symmetry systems. For example, the 
$(110)$
surface of a bcc lattice has a twofold rotational symmetry and two perpendicular mirror planes. For this symmetry class, antiferromagnetic spin spirals have been observed in atomically thin Mn and Cr layers on W(110)~\cite{Bode2007,Yoshida2012,santosNewJ.Phys.2008} and an anharmonic spin spiral or a periodic array of ferromagnetic domain walls was found in two atomic layers of 
Fe on W(110)~\cite{PhysRevLett.103.157201,PhysRevB.78.140403}. In these systems, the strong anisotropy leads to the formation of a single orientational domain with Néel-type rotation enforced by the DMI, inhibiting the formation of 
non-coplanar magnetic states at domain boundaries. 
In addition, the anisotropy of the in-plane directions of the films has prevented the experimental observation of skyrmions in such systems so far, despite intriguing theoretical predictions on their properties~\cite{Hoffmann2017,PhysRevB.95.214422}.

Here, we report on unconventional domain walls and skyrmions in an Fe monolayer on Ta(110). We perform spin-polarized scanning tunneling microscopy (SP-STM) experiments to characterize the spin structures and describe them with spin-model simulations based on parameters obtained from first-principles calculations~\cite{Rozsa2015}. Due to the single-atomic thickness of the magnetic film, the structures are 
governed 
by local magnetic interactions instead of the dipolar interaction. Yet, we identify both 
spin-spiral domain walls containing multiple periods of rotations and chains of non-coplanar merons and antimerons. In extended Fe films, we find that the structure of the domain walls is defined by their crystallographic orientation,  
since the anisotropic surface prefers spin modulation only along the 
$[1\overline{1}0]$ 
direction.  
In samples with coalesced Fe islands, the energy balance is changed  
due to a higher step-edge and 
hole density, and  
the domain walls of the continuous film extend into a continuous spin spiral. Under the application of an out-of-plane  
external magnetic field, this spin spiral transforms to a saturated collinear state 
in which elongated skyrmions emerge. 

\section*{Results and Discussion}\label{sec2}

\subsection*{Experimental Results for an Extended Layer}

\begin{figure}[htb]
\centering
\includegraphics[width=0.6\columnwidth]{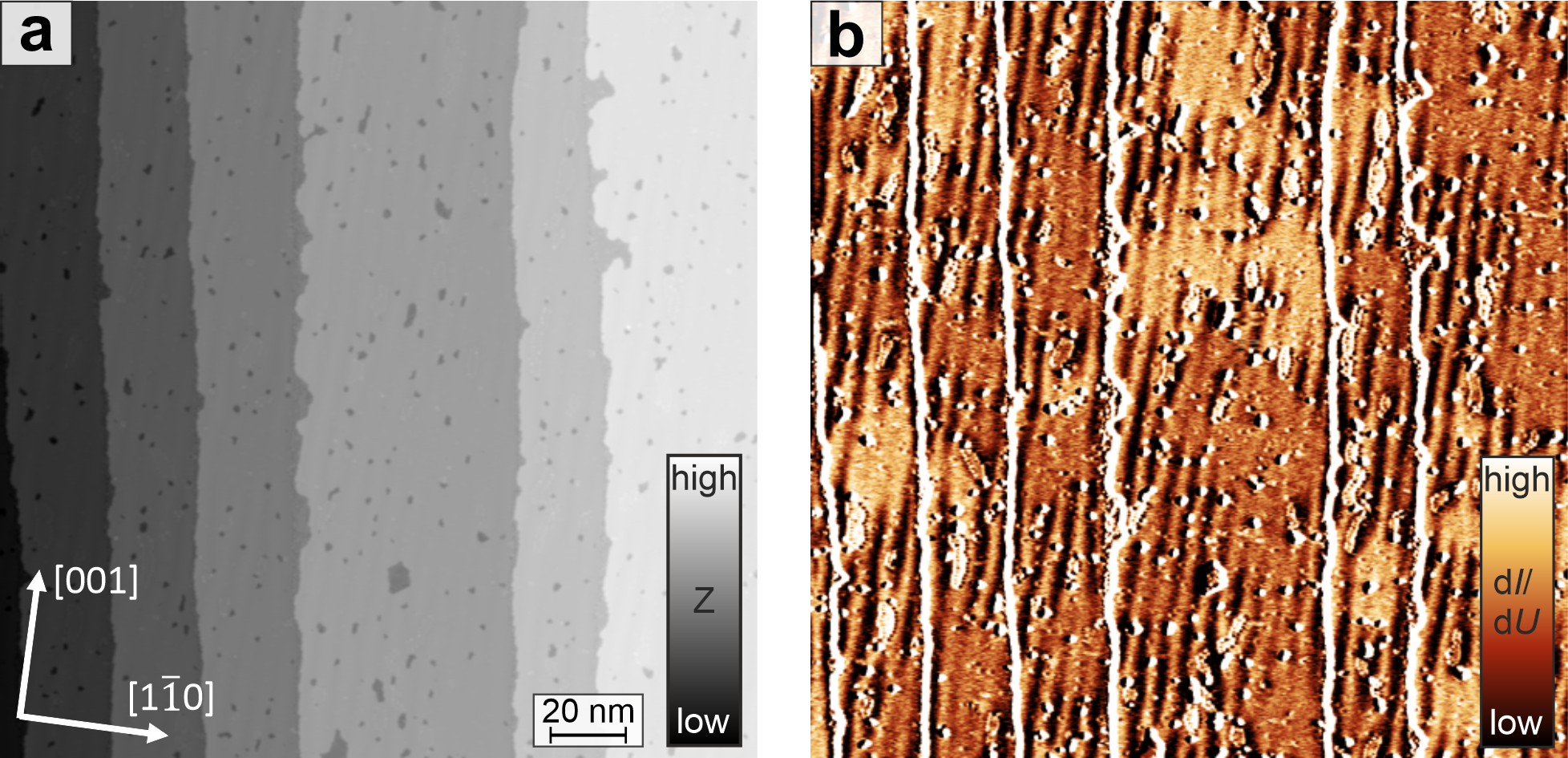}
\caption{\textbf{The morphology and magnetism of the nearly closed pseudomorphic Fe monolayer on Ta(110).}
\textbf{a}, Constant-current STM image of about 0.95 atomic layers of Fe on Ta(110).
\textbf{b}, Simultaneously acquired spin-resolved d$I$/d$U$~map. The 
variation of the signal is 
due to the spin-polarized contribution to the tunnel current and demonstrates a complex magnetic state. 
Measurement parameters: $I = 1$~nA, $U = -100$~mV, $T = 4.2$~K, Cr bulk tip.
}
\label{fig:expIntro}
\end{figure}

We studied the system of an Fe monolayer on the surface of a clean Ta(110) substrate. Figure~\ref{fig:expIntro}a displays the topography of a sample with an Fe coverage of about 0.95 atomic layers and several buried step edges of the Ta substrate. Up to a coverage of a fully closed monolayer, Fe can be grown pseudomorphically on Ta(110) when the substrate is well above room temperature during deposition, while for 
higher coverages or lower substrate temperatures several different Fe reconstructions are formed~\cite{BrueningPhD}. To investigate the magnetic state of the Fe monolayer, we employ SP-STM, where a magnetic tip is used and spin-polarized tunneling between the magnetic sample and the magnetic tip provides information regarding the sample magnetization~\cite{WiesendangerRMP2009,BergmannJPCM2014}. In Fig.~\ref{fig:expIntro}b, we present the spin-resolved differential tunneling conductance d$I/$d$U$ of the sample area displayed in Fig.~\ref{fig:expIntro}a. We find a complex pattern, which can be described as a coexistence of nano-scale stripes along $[001]$ and more extended areas of uniform signal with two different signal intensities. 

\begin{figure}[htb]
\centering
\includegraphics[width=0.5\columnwidth]{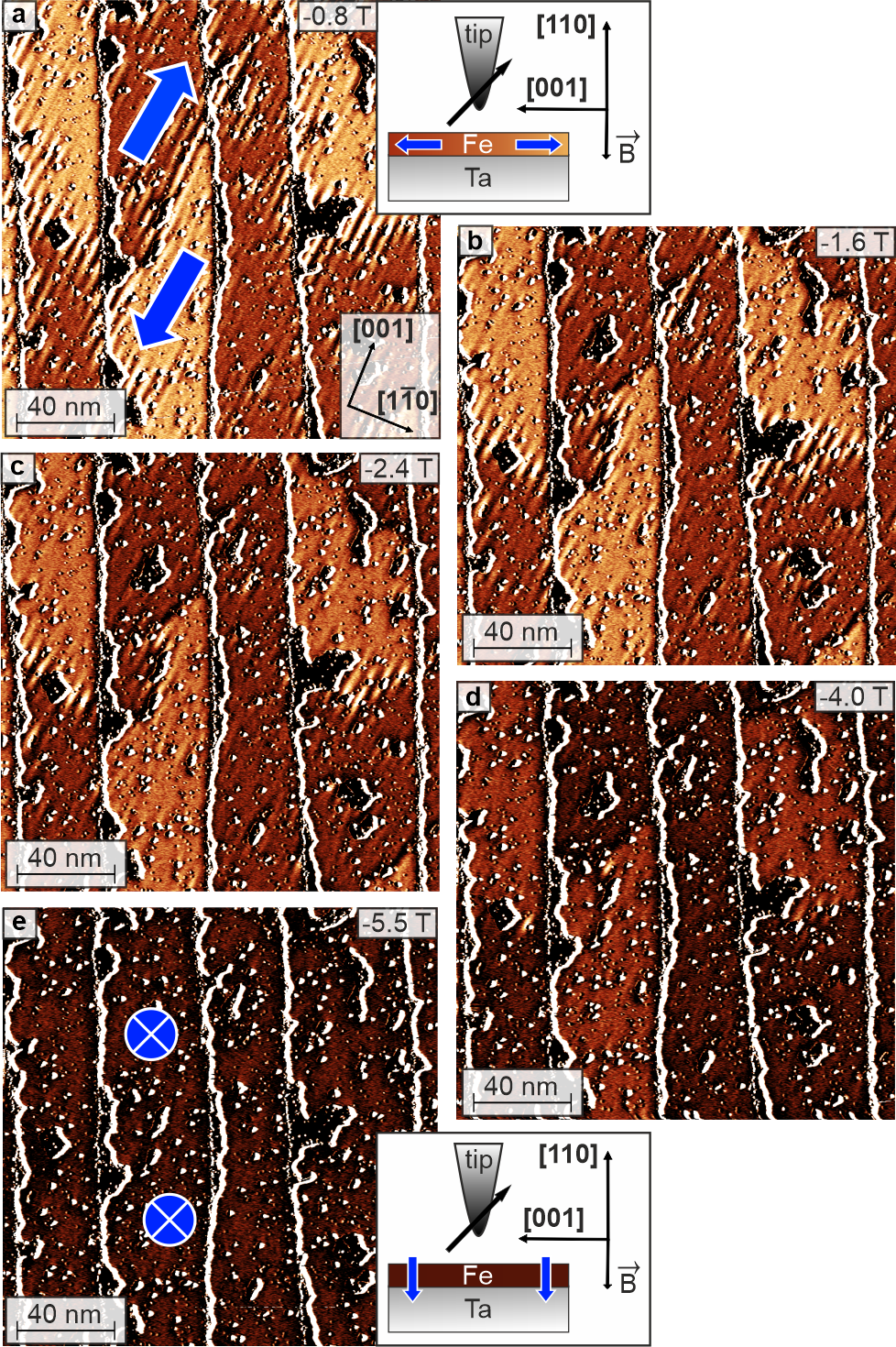}
\caption{\textbf{Field dependence of the magnetic order in the extended Fe monolayer on Ta(110).}
\textbf{a-e}, Spin-resolved d$I$/d$U$~maps of the same area with increasing external magnetic field as indicated; the signal amplitude 
remains roughly constant throughout this measurement series, which indicates that the tip magnetization direction $\mathbf{m_t}$ does not respond to the applied magnetic field, but remains fixed during the entire experiment. The insets in \textbf{a} and \textbf{e} are schematic representations of the experiment at zero field and saturation, respectively. Measurement parameters: $I = 1$~nA, $U = -0.5$~V, $T = 4.2$~K, Cr bulk tip.}
\label{fig:expDomainsInField}
\end{figure}
\noindent

We first focus on the extended uniform areas, where the observed 
two-stage contrast in the differential tunneling conductance is typically a sign of oppositely magnetized ferromagnetic domains. Due to the symmetry of the system, this two-stage contrast could originate from  
ferromagnetic order along any of the high-symmetry directions, i.e.\ $[001]$, $[1\overline{1}0]$, or out-of-plane. We can discriminate between in-plane and out-of-plane magnetization by application of magnetic fields: in the case of out-of-plane magnetic field, we expect the motion of domain walls for out-of-plane ferromagnets, whereas in-plane ferromagnets should gradually rotate their magnetization direction towards the applied magnetic field. Such a series of magnetic-field-dependent measurements is shown in Fig.~\ref{fig:expDomainsInField}a-e. These measurements are performed with a Cr bulk tip which does not respond to applied magnetic fields. At small external magnetic fields (Fig.~\ref{fig:expDomainsInField}a) we observe a strong two-stage contrast for the extended uniform areas. Upon an increase of the external magnetic field, we find that both the brighter area and the darker area show a decrease of the signal intensity until they are not distinguishable anymore at $-5.5$~T (Fig.~\ref{fig:expDomainsInField}e). We conclude that at this magnetic field the extended areas are mostly aligned with the applied out-of-plane magnetic field, as indicated by the two down-pointing arrows in the inset of Fig.~\ref{fig:expDomainsInField}e. Concurrently, this implies that the two-stage contrast at lower magnetic field values originates from the in-plane magnetization components (see inset of Fig.~\ref{fig:expDomainsInField}a). We conclude that at zero magnetic field, the extended areas are magnetized in the film plane. 

\begin{figure*}[htb]
\centering
\includegraphics[width=0.9\textwidth]{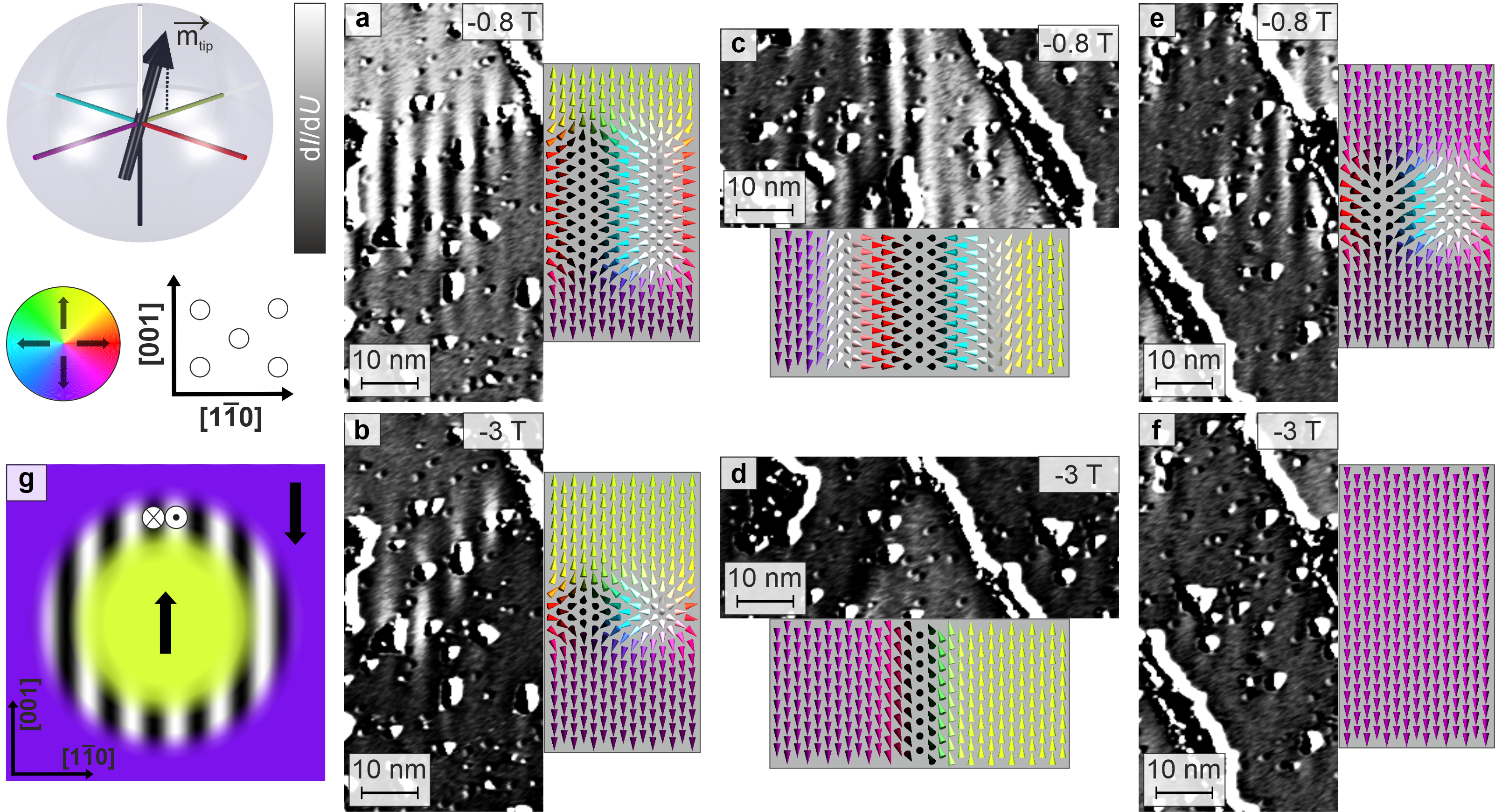}
\caption{
\textbf{Boundaries between domains.}
\textbf{a,b}, 
Spin-resolved d$I$/d$U$ maps of a domain wall between opposite ferromagnetic domains which runs roughly along $[1\overline{1}0]$ at different external magnetic fields as indicated. 
\textbf{c,d}, 
Spin-resolved d$I$/d$U$ maps of a domain wall between opposite ferromagnetic domains which runs roughly along $[001]$ at different external magnetic fields as indicated. 
\textbf{e,f}, 
d$I$/d$U$ maps of a stripe area within the same ferromagnetic domain at different external magnetic fields as indicated. The insets in panels \textbf{a}-\textbf{e} show illustrations of the respective spin configurations; out-of-plane magnetization components are indicated in gray-scale, whereas in-plane components are coloured according to the colour wheel to the left.
\textbf{g}, Sketch of a circular domain wall.
Measurement parameters: $I = 1$~nA, $U = -0.5$~V, $T = 4.2$~K, Cr bulk tip.
}
\label{fig:expDW}
\end{figure*}
\noindent

The more complex SP-STM patterns observed in the Fe monolayer (Fig.~\ref{fig:expDomainsInField}a) can be characterized as alternating bright and dark stripes along $[001]$, with 
a periodicity on the order of 6~nm.
This is reminiscent of a spin spiral with a propagation direction along $[1\overline{1}0]$. 
We find that such spin spiral stripes are always found at 
the transitions between the brighter and darker extended in-plane magnetized ferromagnetic domains, 
i.e., the spin spiral regions act as domain walls. 

Figure~\ref{fig:expDW}a displays a closer view of a characteristic spin-spiral domain wall, where the dividing line between the ferromagnetic domains is roughly along the $[1\overline{1}0]$ direction of the crystal. 
Typically, spin spirals at surfaces are stabilized by DMI and are of cycloidal nature~\cite{Bode2007,BergmannJPCM2014}, i.e., their rotational plane includes the modulation vector $\boldsymbol{q}\parallel[1\overline{1}0]$. 
These spiral stripes may only smoothly connect to the domains on both sides of the wall without an observable distortion, 
if the magnetization direction of the ferromagnetic domains is perpendicular to the spin rotation plane of these cycloidal walls, i.e., along [001], as indicated by the arrows in Fig.~\ref{fig:expDomainsInField}a. 
The inset to Fig.~\ref{fig:expDW}a shows a tentative model of the spin structure: 
the cones indicate the directions of the magnetic moments and their colours correspond to the in-plane magnetization components, while white and black mark opposite out-of-plane magnetization components. The magnetization across the wall can be described as a conical spin spiral with propagation vector along $[1\overline{1}0]$, and a continuous change of the cone angle relative to [001] from $0^\circ$ in the upper domain, via $90^\circ$ in the center of the wall, 
to a cone angle of $180^\circ$ in the lower ferromagnetic domain. The $90^\circ$ cone opening angle in the center of the wall corresponds to a coplanar cycloidal spin spiral. For increasing applied out-of-plane magnetic fields, see Fig.~\ref{fig:expDW}b, the width of the spin-spiral domain wall decreases, i.e., the cone angle across the domain wall changes more rapidly, see inset. 

The transition between the two ferromagnetic in-plane domains exhibits different characteristics when the dividing line is along $[001]$, see Fig.~\ref{fig:expDW}c. Here, from left to right, the spins need to first 
rotate out of the plane to connect smoothly to the cycloidal spin spiral that acts as a domain wall, and then  
rotate back towards [001] on the right side to connect to the right ferromagnetic domain. When the out-of-plane magnetic field is increased, see Fig.~\ref{fig:expDW}d, the domain wall gets thinner by reducing the number of spin-spiral periods. 

We also find areas with spin spiral stripes inside a single ferromagnetic domain, see Fig.~\ref{fig:expDW}e,f. Our tentative model shows how the cone angle varies from the top to the center of the spin spiral area, but then goes back to the initial value in the bottom part of the spin model. Consequently, we find that these spin spiral areas can smoothly vanish when the applied out-of-plane magnetic field is increased. 

We conclude that our system exhibits in-plane ferromagnetic domains that are separated by domain walls consisting of cycloidal spin spirals with a rotational plane perpendicular to the 
magnetization direction in the ferromagnetic domains. Even though the spin-spiral domain walls along the two different high-symmetry directions have been described in slightly different terms, they are of the same type and can be smoothly connected, for instance to form a circular spin-spiral domain wall as in a bubble, as illustrated in Fig.~\ref{fig:expDW}g. This coexistence of ferromagnetic domains and local spin-spiral configurations suggests a delicate balance between the relevant magnetic interaction energies.


\subsection*{Atomistic Spin Model for an Extended Layer}

To understand the origin of these spiral 
domain walls, we theoretically investigate the system based on an atomistic spin model. 
We use the following Hamiltonian to describe the system:
\begin{align}
H=&\frac{1}{2}\sum_{\langle i,j\rangle_{d}}J_{d}\boldsymbol{S}_{i}\boldsymbol{S}_{j}+\frac{1}{2}\sum_{\langle i,j\rangle_{1}}\boldsymbol{D}_{ij,1}\left(\boldsymbol{S}_{i}\times\boldsymbol{S}_{j}\right)\nonumber\\&+\frac{1}{2}\sum_{\langle i,j\rangle_{1}}\left(\Delta J_{1}^{xx}S_{i}^{x}S_{j}^{x}+\Delta J_{1}^{yy}S_{i}^{y}S_{j}^{y}\right)\nonumber\\&+\sum_{i}\left[K^{xx}\left(S_{i}^{x}\right)^{2}+K^{yy}\left(S_{i}^{y}\right)^{2}\right]-\sum_{i}\mu_{\textrm{S}}\boldsymbol{B}\boldsymbol{S}_{i},\label{eq:Ham}
\end{align}
where the $\boldsymbol{S}_{i}$ are unit vectors located at the Fe sites, the $J_{d}$ are isotropic Heisenberg exchange interactions between the nearest ($d=1$) and next-nearest ($d=2$) neighbours, the $\boldsymbol{D}_{ij,1}$ are nearest-neighbour DMI 
vectors, the $\Delta J_{1}^{xx},\Delta J_{1}^{yy},K^{xx}$ and $K^{yy}$ are two-site and single-site anisotropy parameters, $\mu_{\textrm{S}}$ is the magnetic moment and $\boldsymbol{B}$ is the external magnetic field. The $x,y$ and $z$ directions correspond to the $[1\overline{1}0],[001]$ and $[110]$ crystallographic directions, respectively. The two Heisenberg interactions, the different anisotropy coefficients along $x$ and $y$, as well as the two independent components of the 
DM vector capture the anisotropy between the $[1\overline{1}0]$ and $[001]$ directions. For a system with threefold or fourfold rotational symmetry around the out-of-plane direction, only one of each parameter would appear in the model. 

\begin{figure}[!htbp]
\centering
\includegraphics[width=0.5\columnwidth]{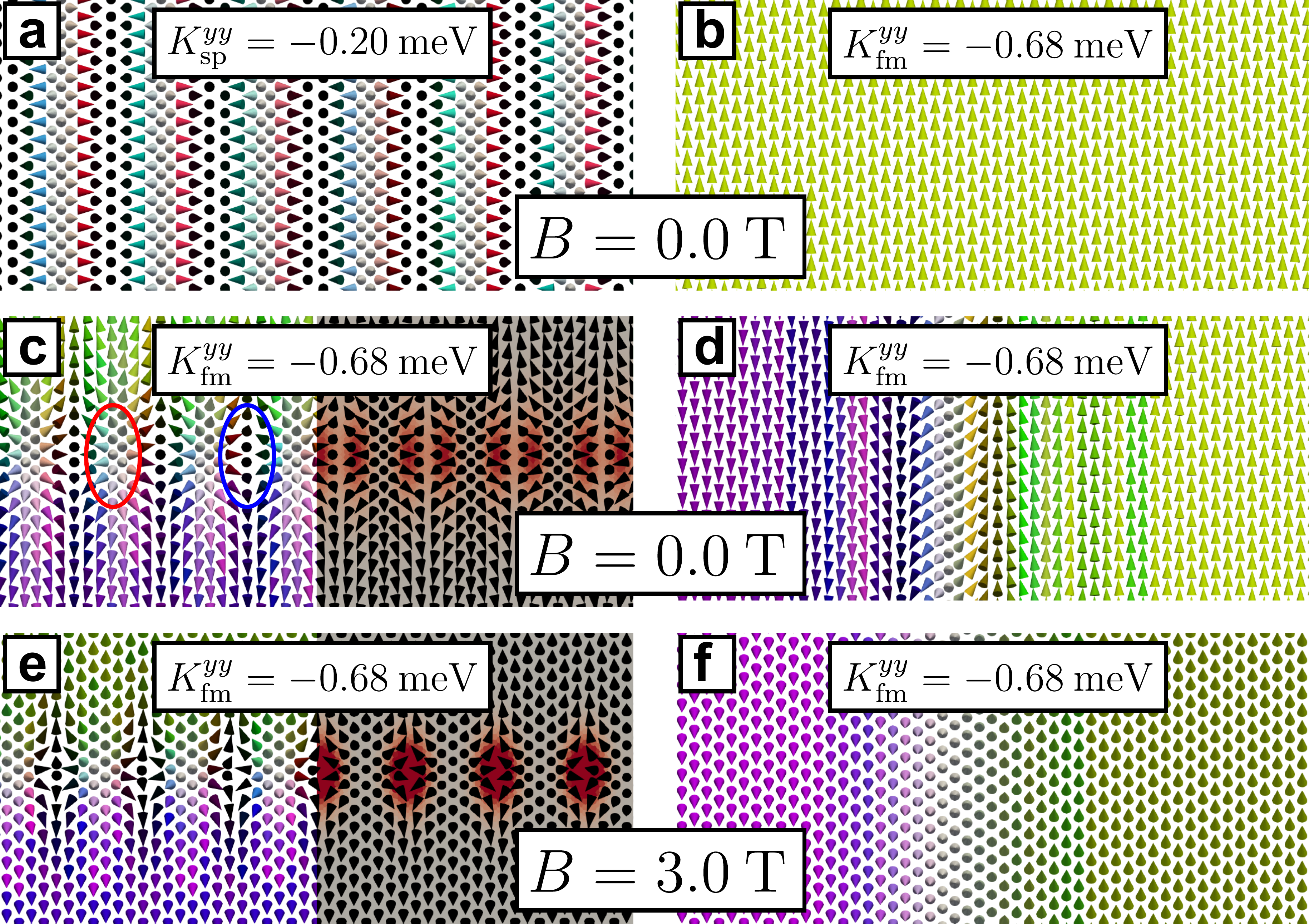}
\caption{\textbf{Ground states and domain walls obtained from spin-dynamics simulations.}
Ground state of the model Hamiltonian in Eq.~\eqref{eq:Ham} for (a) $K^{yy}_{\textrm{sp}}=-0.20$~meV and (b) $K^{yy}_{\textrm{fm}}=-0.68$~meV. Domain walls between two opposite ferromagnetic domains along the $[1\overline{1}0]$ direction are shown at \textbf{c}, $B=0.0$~T and \textbf{e}, $B=3.0$~T out-of-plane fields, and domain walls along the $[001]$ direction are displayed at the same \textbf{d}, $B=0.0$~T and \textbf{f}, $B=3.0$~T field values. Colour indicates the spin directions in three dimensions according to the colour representation in Fig.~\ref{fig:expDW}. Only every ninth atomic spin is shown. Colouring in the background of the right-hand side of \textbf{c} and \textbf{e} shows 
the topological-charge density  
(grey: vanishing, red: positive, blue: negative). Red and blue ovals in \textbf{c} denote a meron and an antimeron, respectively.
}
\label{fig:thDW1}
\end{figure}

The parameter values are listed in Table~\ref{tab:interactions}, together 
with an illustration of the spin model.
These are based on the first-principles calculations in Ref.~\cite{Rozsa2015}, where a cycloidal spin-spiral ground state with approximately 5~nm period along the $[1\overline{1}0]$ direction was found (see Methods), resembling the spin-spiral domain walls in the experiment. The modulation direction of the spin spiral is selected by the interplay between the strong ferromagnetic next-nearest-neighbour interaction $J_{2}$ which suppresses modulations along the $[001]$ direction, and the DMI which prefers rotations along the $[1\overline{1}0]$ direction according to the ratio $D_{1}^{y}/D_{1}^{x}>1/\sqrt{2}$. 
The ground state obtained from spin-dynamics simulations (see Methods) using the parameters in Table~\ref{tab:interactions} is shown in Fig.~\ref{fig:thDW1}a. 

To obtain the in-plane ferromagnetic ground state along the $[001]$ direction, which appears to be the most preferred state for extended films based on the SP-STM data, we increased the magnitude of the anisotropy parameter $K^{yy}$ preferring this direction. 
When increasing the absolute value of $K^{yy}$, the cycloidal spin spiral turns into an elliptic conical spin spiral with the cone axis along the $[001]$ direction, while keeping the modulation direction. For the value of $K^{yy}_{\textrm{fm}}=-0.68$~meV, 
the cone angle closes and the system is ferromagnetic,
see Fig.~\ref{fig:thDW1}b.
Note that in systems with  
threefold or fourfold rotational symmetries a conical spin spiral cannot be the ground state of the  
model in the absence of an external magnetic field~\cite{Banerjee2014}. For the $C_{2\textrm{v}}$ symmetry class, as in the system studied here, this transition through the conical state as a function of anisotropy was previously studied theoretically in Refs.~\cite{Heide2011,Rozsa2016}.

\begin{table}[!htbp]
\parbox{0.48\columnwidth}{
\begin{tabular}{cr}
parameter & value (meV) \\
\hline
$J_{1}$ & -9.25 \\
$J_{2}$ & -19.04 \\
$D_{1}^{x}$ & 1.28 \\
$D_{1}^{y}$ & 2.93 \\
$\Delta J_{1}^{xx}$ & 0.25 \\
$\Delta J_{1}^{yy}$ & 0.21 \\
$K^{xx}$ & 1.29 \\
$K^{yy}_{\textrm{sp}}$ & -0.20 \\
$K^{yy}_{\textrm{fm}}$ & -0.68
\end{tabular}
}
\parbox{0.25\columnwidth}{
\includegraphics[width=0.25\columnwidth]{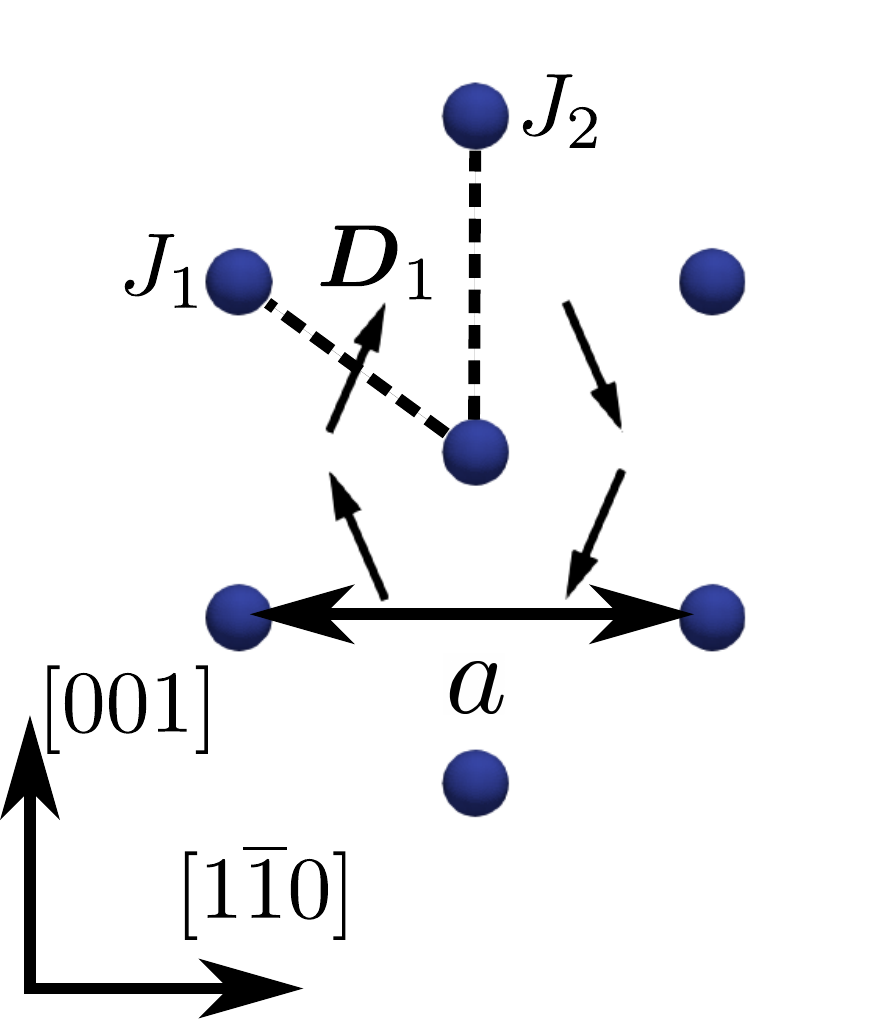}
}
\caption{Parameter values used in the spin model in Eq.~\eqref{eq:Ham}. The ground state is a spin spiral or a ferromagnetic state depending on whether the value $K^{yy}_{\textrm{sp}}$ or $K^{yy}_{\textrm{fm}}$ is used. The DM vectors only have in-plane components in the considered symmetry class. The magnetic moment is $\mu_{\textrm{S}}=2.51~\mu_{\textrm{B}}$. The figure illustrates the interaction coefficients between nearest- and next-nearest neighbours. The components of the DM vector are given for the upper left neighbour as shown in the figure.}
\label{tab:interactions}
\end{table}

\noindent

Domain walls between ferromagnetic domains obtained from the simulations are also shown in Fig.~\ref{fig:thDW1}. If the domains are separated by a wall running along the $[1\overline{1}0]$ direction, a N\'{e}el wall 
would correspond to the simplest structure 
as it avoids the hard axis. 
Instead a spiral modulation is observed between the two domains, with the spins rotating in the $[1\overline{1}0]-[110]$ plane in the middle of the wall, and closing in on the ferromagnetic directions in the two domains by forming a conical structure. The extent of the conical modulation decreases when the magnetic field is increased from Fig.~\ref{fig:thDW1}c to e, but the period of the modulation does not change considerably. 
Inside the domain wall, two alternating spin structures can be observed: merons, in which all spins are pointing away from the central out-of-plane spin, denoted by a red oval in Fig.~\ref{fig:thDW1}c; and antimerons, in which some spins are pointing towards while other spins are pointing away from the center out-of-plane spin, as shown inside the blue oval in Fig.~\ref{fig:thDW1}c.

The topological-charge density inside the domain wall along the $[1\overline{1}0]$ direction is shown on the right-hand side of Fig.~\ref{fig:thDW1}c and e (see Methods). 
The sign of the topological-charge density is determined by whether the in-plane components of the spins form a meron or antimeron 
structure, and whether the out-of-plane component in the center is pointing up or down, i.e., the polarity of the meron.  
Since the meron  
and antimeron  
structures inside the domain wall have opposite polarity, the sign of the topological-charge density does not change along the wall, as can be seen in the figure. 
Note that the magnitude of the topological-charge density is not uniform along the wall already at $B=0.0$~T in Fig.~\ref{fig:thDW1}c. The angles between neighbouring spins are largest when they are close to the $[1\overline{1}0]$ hard axis, 
leading to a local enhancement of the topological-charge density. 
Furthermore, the directions of the DM vectors in the system prefer the formation of merons over antimerons. Therefore, inside the wall the merons and antimerons are slightly extended and compressed, respectively, leading to a larger topological-charge density at the antimerons' location. 

When two opposite ferromagnetic domains are separated by a wall running along the $[001]$ direction, the spins rotate out of the plane to connect the two domains. While in this arrangement a Bloch wall with spins in the $[001]-[110]$ plane would be the geometrically simplest configuration, the spins also acquire a finite component along $[1\overline{1}0]$ which is preferred by the DMI.  
At low magnetic fields in Fig.~\ref{fig:thDW1}d, 
an oscillatory structure resembling multiple periods of the spiral can be observed around the middle of the wall. The width of the wall decreases for increasing field in Fig.~\ref{fig:thDW1}f,  
and the wall vanishes when the magnetic field aligns the domains along the out-of-plane direction at around $B=3.8$~T in the spin model. 
A wall along this direction does not exhibit a net topological charge density since the spins are only modulated along one spatial direction.

These results are in excellent agreement with the spin models deduced based on the experimental data in Fig.~\ref{fig:expDW} and demonstrate how competing interactions can lead to complex magnetic order. This system is very sensitive to changes of the anisotropy, which in turn can vary locally in imperfect films. 


\subsection*{Experimental Results for Connected Islands}

\begin{figure}[htb]
\centering
\includegraphics[width=0.5\columnwidth]{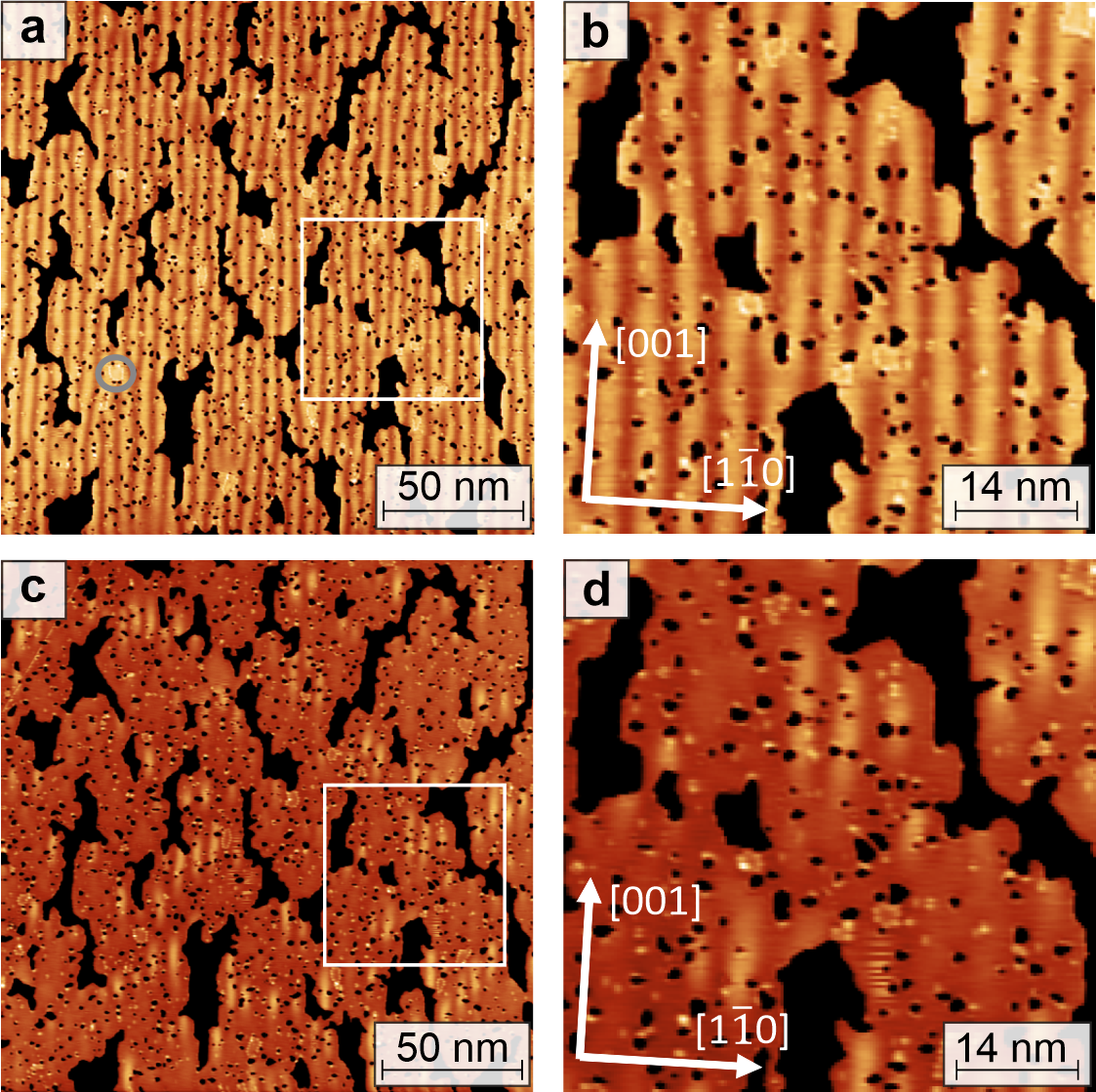}
\caption{\textbf{Magnetic state of the Fe monolayer with holes. }
\textbf{a}, Constant-current SP-STM image of an Fe monolayer with holes.
\textbf{b}, Magnified view of the area indicated by the white box in panel a. 
\textbf{c}, Same as in panel \textbf{a} with an applied out-of-plane magnetic field of $B=2.4$~T.
\textbf{d}, Magnified view of the area indicated by the white box in panel \textbf{c}. 
Measurement parameters: $I = 1$ nA, $U = -0.04$ V, $T = 4.2$ K, Cr bulk tip.
} 
\label{fig:expSS}
\end{figure}

\noindent

We can change the sample morphology when the Fe deposition is performed at slightly lower substrate temperature, which leads to preferential island growth instead of step-flow growth (see Methods). An overview SP-STM constant-current image of such a sample is shown in Fig.~\ref{fig:expSS}a, and we find that the Fe layer has more 
vacancy islands compared to the samples discussed in Figs.~\ref{fig:expIntro}-\ref{fig:expDW}. In the sample shown in Fig.~\ref{fig:expSS}, the spin spiral is present in the entire Fe film, except for the small reconstructed Fe patches visible as bright squares (see, e.g., gray circle), similar to a previous study~\cite{RBarxiv}. The white square indicates the sample area shown again in Fig.~\ref{fig:expSS}b, where one can see that in addition to the vacancy islands also many small holes  
are present in the Fe film. Figures~\ref{fig:expSS}c,d show the respective sample areas imaged in an applied out-of-plane magnetic field of $2.4$~T. These SP-STM measurements show a few bright objects that are elongated along $[001]$ within a homogeneous darker background. This is reminiscent of the typical transition from a spin spiral phase to isolated skyrmions seen for 
systems with three-fold rotational symmetry~\cite{RommingS2013,HsuNC2018}, where the skyrmions are axially symmetric. In contrast to another 
system exhibiting bean-shaped skyrmions~\cite{HsuNN2017}, where 
the deformation of the skyrmions is driven by the underlying reconstruction,
in our system the  
elongated shape is due to the anisotropy of the magnetic interactions and reflects  
the underlying crystal symmetry.


\subsection*{Atomistic Spin Model Including Defects}

\begin{figure}[htb]
\centering
\includegraphics[width=0.5\columnwidth]{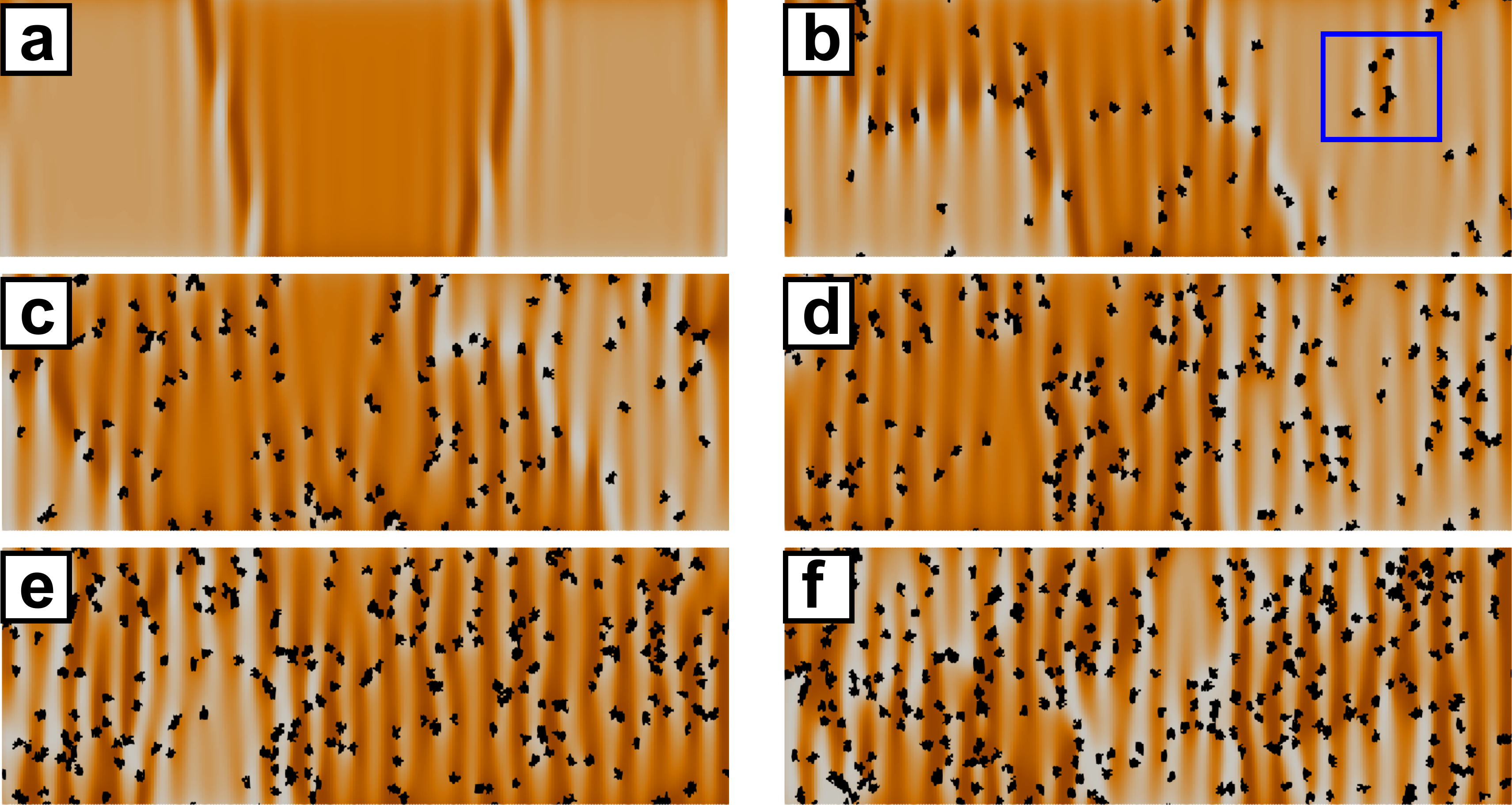}
\caption{\textbf{Spin configurations obtained from the spin model when the anisotropy is changed around the 
holes, for different numbers of holes. 
}
The nominal number of  
holes is 0 in \textbf{a}, 60 in \textbf{b}, 120 in \textbf{c}, 180 in \textbf{d}, 240 in \textbf{e}, and 300 in \textbf{f}, see Methods. The anisotropy was set to $K^{yy}_{\textrm{fm}}$ in the extended areas, 
but it was changed to $K^{yy}_{\textrm{sp}}$ up to two atoms distance around the  
hole sites in panels 
\textbf{b}-\textbf{f}  
to prefer a spin-spiral ground state.  
Holes are denoted by black colour. 
The blue rectangle in \textbf{b} highlights local modulations inside a single domain. 
Colour maps show the projection of the spin directions onto a fixed direction modelling the SP-STM tip, with an angle of $\alpha=20.1^{\circ}$ from the $[110]$ direction in the $[001]-[110]$ plane. The size of the lattice was $256a\times 64\sqrt{2}a$, with free boundary conditions along the edges.
}
\label{fig:thdefectdep}
\end{figure}

We studied the role of edges and 
holes in the formation of modulated spin structures in the simulations by taking multiple effects into account. First, a collinear magnetic  
state  
gets perturbed if some of the neighbouring spins of a site are missing. A typical example of this is the canting of the spins at the edges of out-of-plane magnetized islands caused by the DMI
~\cite{Rohart2013}. Second, the reduced number of neighbours influences the electronic structure, which locally changes the parameters in the spin Hamiltonian itself. For example, first-principles calculations for monatomic Fe chains on Re(0001) found different Heisenberg exchange, DMI 
and anisotropy coefficients at the ends of the chains compared to the middle~\cite{Laszloffy2019}.

The effect of 
holes in the simulations is illustrated in Fig.~\ref{fig:thdefectdep}, where 
the system was cooled down from a paramagnetic state to zero temperature. 
In addition to the spin-spiral domain walls, modulations can be seen at the left and right sample edges in Fig.~\ref{fig:thdefectdep}a, which are parallel to the $[001]$ direction,  
due to the reduced number of neighbours at the free boundaries. In contrast to the 
typical DMI-induced edge tilt, where the spins are confined to the plane perpendicular to the edge, the modulations induced here correspond to a three-dimensional conical rotation of the spins around the ferromagnetic direction. 

For the additional simulations shown in Fig.~\ref{fig:thdefectdep}b-f, we incorporate an increasing number of  
small holes.  
To improve the qualitative agreement with the experiments, we also changed the anisotropy parameter next to the  
holes from $K^{yy}_{\textrm{fm}}$ to $K^{yy}_{\textrm{sp}}$. 
With the increasing number of 
holes in Fig.~\ref{fig:thdefectdep}b-f, the area covered by the modulated structures increases, while the contrast difference between the two domains decreases, indicating an increase in the opening angle of the cone. 
The increase in the areas covered by spirals is not restricted to the domain walls, but local modulations also emerge in the vicinity of defects inside a single domain, see the blue rectangle in Fig.~\ref{fig:thdefectdep}b. These modulations correspond to the stripes observed experimentally inside a single domain in Fig.~\ref{fig:expDW}e. 
Further increasing the number of holes is expected to lead to the disappearance of the two domains 
as the cycloidal spin spiral  
depicted in Fig.~\ref{fig:thDW1}a becomes the magnetic ground state. 

\subsection*{Atomistic Spin Model in External Magnetic Fields}

\begin{figure}[htb]
\centering
\includegraphics[width=0.45\columnwidth]{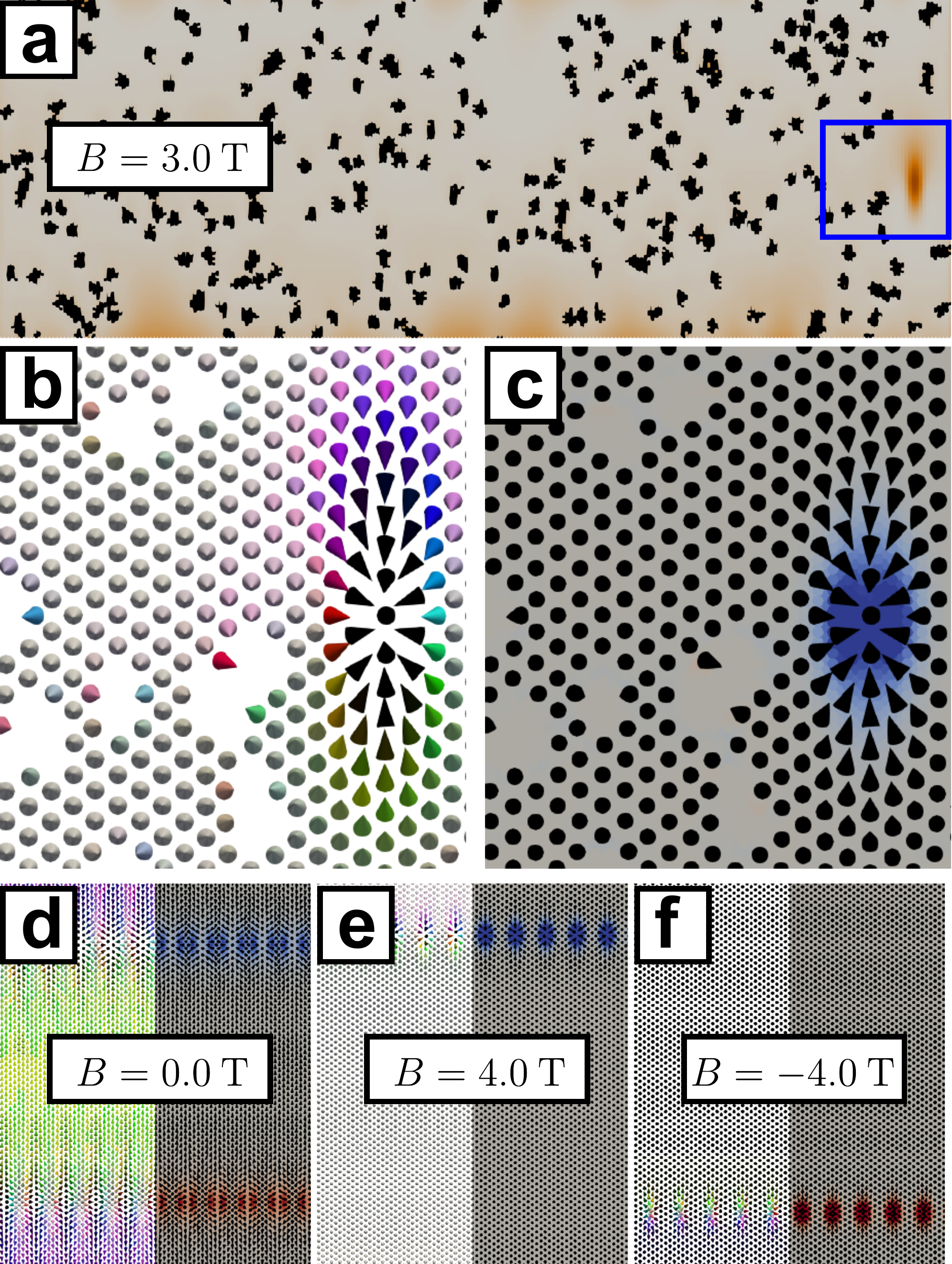}
\caption{\textbf{Elongated skyrmions in an applied magnetic field.}
\textbf{a}, The same simulation cell as in Fig.~\ref{fig:thdefectdep}f after applying an out-of-plane field of $B=3.0$~T. \textbf{b,c} Visualization of the spin direction and topological charge density of the elongated skyrmion from the area highlighted by the blue rectangle in \textbf{a} (colour-coded as in Fig.~\ref{fig:thDW1}).
\textbf{d}, A pair of domain walls along the $[1\overline{1}0]$ direction at $B=0.0$~T. After applying an external field of \textbf{e}, $B=4.0$~T or \textbf{f}, $B=-4.0$~T, one of the domain walls vanishes while the other transforms into a chain of skyrmions. Right sides of the panels show the topological-charge density. Only the direction of every ninth spin is shown.}
\label{fig:Field_dependence_skyr_sim}
\end{figure}

In an external magnetic field, the strongly modulated structure shown in Fig.~\ref{fig:thdefectdep}f changes, and at $B=3.0$~T most regions are aligned along the field direction, see Fig.~\ref{fig:Field_dependence_skyr_sim}a. 
However, modulated magnetic objects occasionally remain, as indicated by the blue rectangle, similar to what has been observed in the experiments (Fig.~\ref{fig:expSS}c,d). The spin configuration of such a structure is displayed in panels b and c, confirming that it is an elongated skyrmion with a finite topological charge. 

We also investigated in more detail the response to external magnetic fields for two different (head-to-head and tail-to-tail) $[1\overline{1}0]$ domain walls separating the two opposite ferromagnetic domains in extended Fe monolayer films, see Fig.~\ref{fig:Field_dependence_skyr_sim}d. In zero magnetic field their topological charge is opposite, with a localization of the topological-charge density at the antimerons due to the DM interaction. The asymmetry between the shape of merons and antimerons also gives rise to a net out-of-plane magnetization in the domain wall: it points along the core direction of merons, which is upwards for the domain orientation of the lower domain wall in Fig.~\ref{fig:Field_dependence_skyr_sim}d, and opposite for the upper domain wall. 
This opposite net moment of the two domain walls is also reflected in their stability in applied magnetic fields, as seen in Fig.~\ref{fig:Field_dependence_skyr_sim}e and f. 
Close inspection of the spin structure shows that due to a rotation of the ferromagnetic domains towards the positive applied field in Fig.~\ref{fig:Field_dependence_skyr_sim}e, the merons of the upper domain wall are transformed into skyrmions, whereas the antimerons smoothly vanish and merge with the saturated background. For the lower domain wall, the merons can merge with the background, whereas antimerons are continuously transformed into antiskyrmions. Due to the DM interaction, the critical magnetic field for the annihilation of antiskyrmions is lower, and only the energetically preferred skyrmions in the upper wall persist up to around $6.8$~T. For negative field directions in Fig.~\ref{fig:Field_dependence_skyr_sim}f, the situation is reversed, and a chain of skyrmions survives in the lower domain wall.

\section*{Conclusion}\label{sec13}
In summary, we  
demonstrated the presence of unconventional
topologically distinguishable types of
domain walls and elongated skyrmions in an Fe monolayer on Ta(110). Due to a subtle balance of different magnetic energy terms, non-coplanar domain walls emerge in the in-plane ferromagnetic ground state of an extended atomic layer, which can be described in terms of conical modulations. 
The internal spin structure of the domain walls is determined by their crystallographic orientation. Whereas domain walls along $[001]$ are topologically trivial, those along $[1\overline{1}0$] can be viewed as a chain of merons and antimerons and have a net topological charge of 1 per 6 nm wall length. This domain wall is reminiscent of cross tie walls in ferromagnetic films, however, in 
contrast to ferromagnets with strong dipolar interactions, where layer thickness and shape play a decisive role, 
here the spin structure in the domain wall is governed by the interplay of exchange, anisotropy, and DM interaction. Domain walls in spin spirals may also constitute of chains of merons depending on the angle between the modulation vectors in the two domains~\cite{PhysRevLett.108.107203}, while in the ferromagnetic system here the crystallographic orientation of the domain wall is relevant for the morphology of domain walls instead. 

For sample morphologies with a high number of 
vacancies and
holes in the film, we found that the domain walls merge to form a  
spin spiral.  
We understand this effect as a result of the edge effects caused by the DMI and the 
modification of the 
magnetocrystalline anisotropy energies in the vicinity of edges. 
Interestingly, both the extended layers with ferromagnetic domains and topological domain walls, and the connected islands with spin spirals, lead to elongated skyrmions in applied magnetic fields.

Our study is expected to motivate further research into low-symmetry magnetic systems which can be achieved by the proper surface orientation, growing wedge-shaped samples or by strain engineering in two-dimensional materials. The extended parameter space of local interactions in such structures may lead to the discovery of further exotic spin configurations, possibly with unusual current-induced motion 
which may be taken advantage of in nanoscale data-storage and logic devices. Additionally, the meron-antimeron domain wall constitutes a non-coplanar one-dimensional magnetic defect in direct vicinity to superconducting Ta, which is expected to feature novel emerging properties useful in future quantum technologies. 

\section*{Acknowledgements}
K.v.B.\ gratefully acknowledges financial support from the Deutsche Forschungsgemeinschaft (DFG, German Research Foundation) via project no.\ 402843438. R.B. and R.W. acknowledge funding by the European Research Concil (ERC Advanced Grant No. 786020). R.L.C.\ acknowledges financial support
from the Deutsche Forschungsgemeinschaft (DFG, German Research Foundation) via grant no. 459025680. L.R. gratefully acknowledges funding by the National Research, Development, and Innovation Office (NRDI) of Hungary under Project Nos. K131938 and FK142601, by the Ministry of Culture and Innovation and the National Research, Development and Innovation Office within the Quantum Information National Laboratory of Hungary (Grant No. 2022-2.1.1-NL-2022-00004), and by the Hungarian Academy of Sciences via a J\'{a}nos Bolyai Research Grant (Grant No. BO/00178/23/11).

\section*{Author contributions}
R.B.\ performed the measurements. R.B., K.v.B., A.K. and R.L.C.\ analyzed the experimental data. L.R.\ performed the simulations and calculations. K.v.B.\ and L.R.\ wrote the manuscript. R.B., L.R., R.L.C., A.K., R.W., and K.v.B.\ discussed the results and the manuscript.

\section*{Methods}

\subsection*{Experimental details}
\textbf{Sample preparation} \\
To obtain a clean Ta(110) surface, the single crystal was flashed by electron-beam heating up to $\sim$2200°C multiple times for 60 seconds. The Fe was evaporated from a rod and deposited on the Ta crystal by physical vapor deposition under ultra-high vacuum conditions with a base pressure of $\sim1.0\times10^{-10}$~mbar. The different sample morphologies were obtained by varying the substrate temperature during the Fe deposition, where samples with a lower substrate temperature show a comparably large number of vacancy islands (as in Fig.~\ref{fig:expSS}), and samples with a higher substrate temperature show step-flow growth with only few defects inside the Fe monolayer.

\textbf{SP-STM} \\
The samples were transferred \emph{in-situ} to a home-built STM system operated at 4.2~K. All SP-STM measurements presented in this work were performed using a bulk Cr tip~\cite{WiesendangerRMP2009,BergmannJPCM2014}. The d$I$/d$U$ measurements were performed using a lock-in technique by adding a small modulation voltage $U_{\textrm{mod}}$ on the order of $10\%$ of the sample bias voltage with a frequency of 4777 Hz to the bias voltage. 


\subsection*{Theoretical calculations}

\textbf{Spin-model parameters} \\
The spin-model parameters in Ref.~\cite{Rozsa2015} were determined from first-principles calculations based on the screened Korringa--Kohn--Rostoker method~\cite{Szunyogh1994,Zeller1995,Szunyogh1995}. It was found there that the magnetic ground state strongly depends on the calculation parameters, particularly on the relaxation, i.e., the distance between the Fe atoms and the Ta(110) substrate. Since the interlayer distance cannot be determined experimentally, we chose the relaxation value of 13.5\% which results in a cycloidal spin spiral ground state with a period and propagation length similar to that observed in the spin-spiral domain walls or the spin-spiral phase in the experiments. The energy dispersions obtained from the original Hamiltonian containing interactions with 562 neighbours in Ref.~\cite{Rozsa2015} were fitted to the simplified model in Eq.~\eqref{eq:Ham} with a total of 6 nearest and next-nearest neighbours. The distinction between two-site and single-site anisotropy terms from the first-principles calculations was preserved in the simplified model. It was found in the first-principles simulations that the modulation period and the rotational plane changes with the relaxation, but the preferred modulation direction along $[1\overline{1}0]$ is robust against the calculation details. The calculations also indicated that the easy axis is reoriented from the out-of-plane $[110]$ to the $[001]$ direction with increasing relaxation, and the change in the anisotropy is of similar magnitude as the difference between the values $K^{yy}_{\textrm{sp}}=-0.20$~meV and $K^{yy}_{\textrm{fm}}=-0.68$~meV, which were used to describe Fe atoms with different atomic environments in this work.

The preferred modulation direction in the plane may be quantitatively examined by introducing the micromagnetic parameters~\cite{Rozsa2017}
\begin{align}
\mathcal{J}_{[1\overline{1}0]}&=-\frac{1}{t}\frac{\sqrt{2}}{2}J_{1},\\
\mathcal{J}_{[001]}&=-\frac{1}{t}\frac{\sqrt{2}}{4}\left(J_{1}+2J_{2}\right),\\
\mathcal{D}_{[1\overline{1}0]}&=\frac{1}{at}2\sqrt{2}D_{1}^{y},\\
\mathcal{D}_{[001]}&=\frac{1}{at}2D_{1}^{x},\\
\end{align}
where $a$ is the distance between consecutive atoms along the $[1\overline{1}0]$ direction and $t$ is a nominal layer thickness. The energy gain per unit volume from forming a spin spiral may be estimated as $\Delta e_{\alpha}=\mathcal{D}^{2}_{\alpha}/\left(2\mathcal{J}_{\alpha}\right)$ for cycloidal spin spirals rotating in the $\alpha-z$ plane with wave vectors along the $\alpha=[1\overline{1}0],[001]$ directions. The anisotropy of the  
DMI and the strong next-nearest-neighbour exchange interaction results in $\Delta e_{[1\overline{1}0]}\approx 25\Delta e_{[001]}$, leading to a strong preference for forming the spin spiral along the $[1\overline{1}0]$ direction. The energy gain from the DMI  
is balanced by the anisotropy terms preferring a collinear alignment, leading to a coplanar cycloidal spin spiral, non-coplanar conical spin spiral or ferromagnetic ground state depending on the value of $K^{yy}$. 

We did not directly include the dipole-dipole interaction in the calculations. Its effect may be estimated for the considered atomic  
layer by a change in the on-site anisotropy term. When summed up over 
the infinite two-dimensional layer (see Ref.~\cite{Szunyogh1995} for the method), the dipolar interactions favour an in-plane orientation by about $\Delta K^{yy}_{\textrm{dip}}\approx-0.10$~meV and $\Delta K^{xx}_{\textrm{dip}}\approx -0.11$~meV, which is smaller than the difference between $K^{yy}_{\textrm{sp}}$ and $K^{yy}_{\textrm{fm}}$ used to transform between the spin spiral and ferromagnetic ground states. This supports the assessment that considering local modulations in the electronic structure are necessary for explaining the change in the spin structure in the vicinity of 
holes.

\textbf{Spin-dynamics simulations} \\
The spin configurations were obtained by numerically solving the stochastic Landau--Lifshitz--Gilbert equation~\cite{Landau1935,Gilbert2004,Brown1963} based on the spin model. The system was initialized in a completely disordered state at a relatively high temperature of $T=160$~K, then gradually cooled down to zero temperature and relaxed to a local energy minimum. The boundary conditions were fixed along the positive and negative $[001]$ directions when simulating the domain walls in Fig.~\ref{fig:thDW1}c-e, and they were kept free when finding the ground state and simulating the  
holes in Fig.~\ref{fig:thDW1}a and b, Fig.~\ref{fig:thdefectdep} and Fig.~\ref{fig:Field_dependence_skyr_sim}.

Holes were created by starting from a site selected uniformly in the atomically perfect structure, then randomly adding neighbouring lattice sites until a target hole size was reached. The probability of including a site in the hole at each step was proportional to its number of neighbours already included in the hole, reaching probability $1$ when all six neighbours are part of the hole. The sizes of 
holes were uniformly randomized between a minimal and maximal value (20-30 sites in Fig.~\ref{fig:thdefectdep}), while the number of holes was nominally fixed, but two or more holes may have possibly merged during the procedure.

\textbf{Calculation of the topological-charge density} \\
The topological-charge density at site $i$ is defined as
\begin{align}
q_{i}=\frac{1}{6\pi}\sum_{\left<i,j,k\right>}\arctan{\frac{\boldsymbol{S}_{i}\left(\boldsymbol{S}_{j}\times\boldsymbol{S}_{k}\right)}{1+\boldsymbol{S}_{i}\boldsymbol{S}_{j}+\boldsymbol{S}_{j}\boldsymbol{S}_{k}+\boldsymbol{S}_{k}\boldsymbol{S}_{i}}},
\end{align}
where the summation runs over the six nearest-neighbour triangles $\boldsymbol{S}_{i}$ forms a part of, and the sites are indexed as $\left<i,j,k\right>$ in a counterclockwise order in each triangle. This formula has been proposed in Ref.~\cite{Berg1981} to calculate the topological-charge density or solid angle on the surface of the sphere in a lattice model accurately.

\end{document}